\documentclass{article}
\usepackage{spconf,amsmath,graphicx,psfrag}
\usepackage{bm}
\usepackage{amsfonts}
\usepackage{amsmath}
\usepackage{subcaption}

\title{MASV: Speaker Verification with Global and Local Context Mamba}
\name{Yang Liu, Li Wan, Yiteng Huang, Ming Sun, Yangyang Shi, Florian Metze}
\address{Meta Platforms, USA \\
\{yangliuai, wwanli, yah, sunming425, yyshi, fmetze\}@meta.com}

\begin{document}
%
\maketitle

\begin{abstract}

Deep learning models like Convolutional Neural Networks and transformers have shown impressive capabilities in speech verification, gaining considerable attention in the research community. However, CNN-based approaches struggle with modeling long-sequence audio effectively, resulting in suboptimal verification performance. On the other hand, transformer-based methods are often hindered by high computational demands, limiting their practicality. This paper presents the MASV model, a novel architecture that integrates the Mamba module into the ECAPA-TDNN framework. By introducing the Local Context Bidirectional Mamba and Tri-Mamba block, the model effectively captures both global and local context within audio sequences. Experimental results demonstrate that the MASV model substantially enhances verification performance, surpassing existing models in both accuracy and efficiency.

\end{abstract}

\section{Introduction}

Speech verification is essential for validating the authenticity and reliability of speech inputs, particularly in edge devices like virtual reality (VR) systems and smart glasses, where high security and privacy are critical. In such applications, the demand for both accurate and efficient verification is paramount. Recent advances in deep learning, especially using architectures like Convolutional Neural Networks (CNNs) and Transformers, have driven substantial progress, setting new standards for accuracy and efficiency in this field.

Despite recent advancements, current models still face significant challenges that limit their performance. Studies on CNN-based methods \cite{desplanques2020ecapa} demonstrate that while these models are effective at capturing local features, they struggle to model the full context of long-sequence audio signals because of their limited receptive fields. This limitation often results in suboptimal performance in complex speech verification tasks. In contrast, Transformer-based models \cite{zhao2023pcf} are adept at modeling long-range dependencies but are hindered by high computational complexity. The quadratic scaling of these models with sequence length reduces their practicality for real-time applications.

To overcome these challenges, recent research has explored large pre-trained models and advanced pooling mechanisms. For example, Attentive Statistics Pooling (ASP) and Progressive Channel Fusion (PCF) have proven effective in enhancing the robustness and accuracy of speaker embeddings \cite{liu2024golden, wang2023vot}. Additionally, the development of innovative loss functions, such as AAM-Softmax and its variants, has improved the discriminative power of embeddings, ensuring better separation of intra-class and inter-class features \cite{al2023real, khan2023battling}.

State-space models (SSMs) \cite{gu2021efficiently, gu2021combining} have recently emerged as a promising approach to overcoming these limitations. SSMs establish long-range dependencies with linear computational complexity, providing an efficient solution for processing long sequences. Models like Mamba \cite{gu2023mamba}, which incorporate SSMs, have demonstrated their effectiveness across domains such as natural language processing \cite{pioro2024moe, yang2024clinicalmamba} and vision tasks \cite{gu2023mamba, zhu2024vision}. However, their potential in the field of speech verification has not been thoroughly explored.

In this paper, we propose the MASV model, a novel architecture that incorporates the Mamba module within the ECAPA-TDNN framework for speaker verification. The primary innovation of this model is the introduction of the Local Context Bidirectional Mamba (LCB-Mamba) and the Tri-Mamba block, which together allow the model to capture both global and local contextual information in audio sequences. The LCB-Mamba improves performance in streaming applications by using local context without depending on future information, which offers a significant advantage over traditional bidirectional models. The Tri-Mamba block further enhances the model’s capability by efficiently integrating both local and global context, surpassing the performance of conventional convolutional layers. This architecture overcomes the limitations of CNNs and Transformers, providing a more efficient and accurate solution for real-time speaker verification. Our experimental results show that MASV outperforms existing models, achieving notable improvements in both verification accuracy and computational efficiency.

\section{Background:  Mamba}

\begin{figure*}[ht]
    \centering
    \begin{subfigure}[b]{\textwidth}
        \centering
        \includegraphics[width=\textwidth]{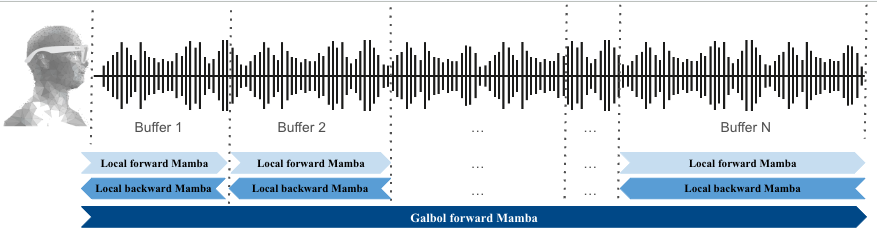}
         \caption{}
        \label{fig:image1}
    \end{subfigure}

    \begin{subfigure}[b]{0.3\textwidth}
        \centering
        \includegraphics[width=\textwidth]{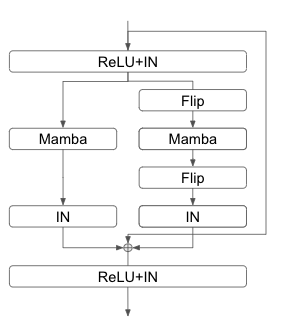}
        \caption{}
        \label{fig:image4}
    \end{subfigure}
    \hfill
    \begin{subfigure}[b]{0.3\textwidth}
        \centering
        \includegraphics[width=0.8\textwidth]{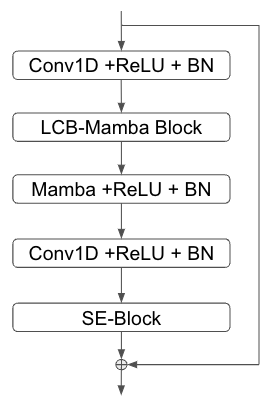}
        \caption{}
        \label{fig:image3}
    \end{subfigure}
    \hfill
    \begin{subfigure}[b]{0.3\textwidth}
        \centering
        \includegraphics[width=0.8\textwidth]{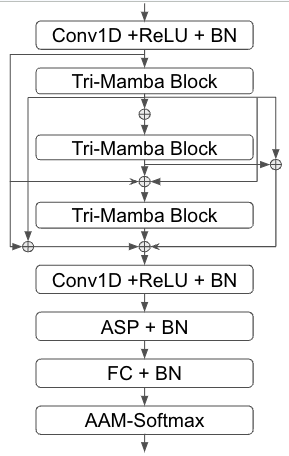}
        \caption{}
        \label{fig:image2}
    \end{subfigure}
    
    \caption{MASV Pipeline and Architectures. (a) Streaming application pipeline where each audio buffer is processed by two independent Mamba layers for local context and a global Mamba layer for accumulated context. (b) Architecture of the LCB-Mamba Block.
(c) Architecture of the Tri-Mamba Block, including the SE (Squeeze-and-Excitation) layer.
(d) Full MASV architecture featuring the ASP (Attentive Statistics Pooling) layer.}
    \label{fig:example}
\end{figure*}

State Space Models (SSMs)  are designed to map a 1-dimensional function or sequence $x(t) \mapsto y(t) \in \mathbb{R} $ and can be represented through a linear Ordinary Differential Equation (ODE):

\begin{align}
x'(t) &= \mathbf{A}x(t) + \mathbf{B}u(t), \\
y(t) &= \mathbf{C}x(t),
\end{align}
where  $\mathbf{A} \in \mathbb{R}^{N \times N}$ is the state matrix, $\mathbf{B}$, $\mathbf{C} \in \mathbb{R}^N$ are parameters and $x(t) \in \mathbb{R}^N$ represents the implicit latent state. The Structured State Space Sequence Models (S4) \cite{gu2021efficiently} enhance basic SSMs by structuring the state matrix $\mathbf{A}$  and implementing an efficient algorithm. The state matrix is initialised using the High-Order Polynomial Projection Operator (HIPPO) \cite{gu2020hippo}, allowing for deep sequence models with robust and efficient long-range reasoning capabilities. This approach has outperformed Transformers \cite{vaswani2017attention} on the Long Range Arena Benchmark \cite{tay2020long}.

The Mamba model \cite{gu2023mamba} extends S4 for discrete data, such as text and genome sequences, with two primary innovations. First, it employs an input-dependent selection mechanism that dynamically adjusts the S4 parameters based on input data, differing from traditional time- and input-invariant S4. Second, Mamba introduces a hardware-aware algorithm that scales linearly with sequence length, enhancing computational efficiency on modern hardware. By integrating S4 blocks with linear layers, Mamba achieves superior performance across various long-sequence tasks, including text, speech, and genomics.

\section{Proposed method}

The proposed MASV model is built upon the ECAPA-TDNN framework, a state-of-the-art approach in speaker verification. Figure 1 presents the overall pipeline of the MASV model, highlighting components designed to address the limitations of traditional CNNs and Transformers in real-time applications. Specifically, we replace the Res2 Dilated CNN with the LCB-Mamba block to better handle local context within audio sequences. In addition, the Tri-Mamba block is introduced to effectively integrate both local and global context, and full-scale skip connections are implemented to facilitate the flow of information across different layers.

\subsection{Local Context Bidirectional Mamba}
In streaming speaker verification, audio data is received in segments known as audio buffers, which can have varying lengths. Traditional bidirectional models typically rely on future context for sequence modeling, which is not feasible in real-time applications where data arrives sequentially. To address this limitation, we propose the Local Context Bidirectional Mamba (LCB-Mamba) block, designed to incorporate local context information within each buffer without depending on future data.

The LCB-Mamba block enhances the model’s performance by integrating local context directly into the state computation. For the local forward Mamba, the state $\overrightarrow{\mathbf{A}}x(t)$ at time step $t$ depends not only on the current input $x(t)$ and the previous hidden state $\overrightarrow{\mathbf{A}}x(t-1)$, but also on a local context window of inputs:
\begin{equation}
\overrightarrow{\mathbf{A}}x(t) = \text{Mamaba}(\overrightarrow{\mathbf{A}}x(t-1), \overrightarrow{\mathbf{c}})
\end{equation}
where $\overrightarrow{\mathbf{c}}$ represents the forward local context in buffer. Similarly, the hidden state $\overleftarrow{\mathbf{A}}x(t)$ is computed by:
\begin{equation}
\overleftarrow{\mathbf{A}}x(t) = \text{Mamaba}(\overleftarrow{\mathbf{A}}_{t+1}, \overleftarrow{\mathbf{c}})
\end{equation}

The final state representation is obtained by concatenating the forward and backward states:
\(\mathbf{A}x(t) = [\overrightarrow{\mathbf{A}}x(t); \overleftarrow{\mathbf{A}}x(t)]\). The LCB-Mamba architecture, as shown in Fig. \ref{fig:image4}, begins with the input passing through a Rectified Linear Unit (ReLU) activation function combined with Instance Normalization (IN). The network then branches into two parallel paths: the forward path consists of a Mamba layer followed by an IN layer, while the backward path includes two Flip layers interspersed with Mamba layers and IN. The outputs from both paths, along with the initial input, are merged and processed through another ReLU layer with IN.

By effectively incorporating local context information, the LCB-Mamba architecture is optimized to capture local dependencies within the buffer, particularly in streaming applications where future context is not available. This design ensures robustness and efficiency in real-time processing scenarios that require high adaptability.

\subsection{Tri-Mamba Block}
While the LCB-Mamba block effectively captures local context, it lacks a mechanism to connect different audio buffers to form a more comprehensive global context. To address this limitation, we propose the Tri-Mamba block, which integrates both local and global context information more efficiently.

As illustrated in Fig. \ref{fig:image3}, the Tri-Mamba block replaces the SE-Res2Block in the original ECAPA-TDNN architecture. The block begins with a Conv1D layer, followed by ReLU activation and Batch Normalization (BN), which extracts essential features from the input. The output then passes through the LCB-Mamba block, which captures bidirectional local context using the mechanisms. To further enhance feature extraction, an additional Mamba block is incorporated with ReLU activation and BN to model more complex patterns. This is followed by another Conv1D layer with ReLU and BN to refine the extracted features. Finally, a Squeeze-and-Excitation (SE) block recalibrates the feature representation by adjusting the channel-wise feature responses.

The combination of these components in the Tri-Mamba block allows for a more comprehensive modeling of audio sequences, effectively leveraging both local and global context information. This design enhances the overall performance of the MASV model, particularly in scenarios requiring real-time speaker verification.

\subsection{MASV with full-scale skip}

To improve information flow and feature retention, we introduce full-scale skip connections in the MASV model. These connections perform element-wise additions between input features and outputs of corresponding layers (Fig. \ref{fig:image2}), preserving both low-level and high-level features. This design mitigates issues like gradient vanishing in deep networks, enhancing training efficiency and convergence.

By combining detailed local information with global context, full-scale skip connections improve accuracy and computational efficiency, making the MASV model robust for real-time speaker verification.

\section{Experiments}

\subsection{Dataset}
We use a geographically filtered private development set as our dataset, consisting of 5,182,021 utterances from 28,622 speakers. This dataset was recorded in a soundproof studio to ensure controlled noise conditions and high-quality audio samples (48 kHz). Our approach addresses the limitations of VoxCeleb2, including noise contamination, uncontrolled recording environments, and privacy concerns, by providing consistent data quality and geographically restricting participation. The input features are 80-dimensional log-Mel filter bank coefficients, normalised via cepstral mean normalisation without voice activity detection. 

\subsection{Model}
Our baseline model is implemented with 512 and 1024 channels, while the proposed MASV (Mamba-based Speaker Verification) model adopts the same configuration with additional enhancements. The MFA module output channels are fixed at 1536 to maintain consistency. 
For comparison, ResNet18, ResNet34 architectures and transformer-based PCF-ECAPA \cite{zhao2023pcf}, are also evaluated using the same attentive statistics pooling mechanism. Training is performed using the Adam optimizer with a cyclical learning rate strategy over 100k-step cycles, varying from 1e-8 to 1e-3, and a weight decay of 5e-5. We employ a batch size of 256 and the Circle loss function with \(m = 0.35\) and \(s = 60\) to enforce stronger constraints on speaker embeddings. All models are trained under these same conditions.

\begin{table}[tp]
\centering
\caption{Performance on Evaluation Sets}
\begin{tabular}{|c|c|c|}
\hline
Model & EER & DCF  \\
\hline
ResNet18  & 1.610 & 0.1989  \\
ResNet34  & 1.264 & 0.1341  \\
ECAPA (C=512) & 1.158 & 0.1221 \\
ECAPA (C=1024)  & 0.956 & 0.1266  \\
PCF-ECAPA (C=512)  & 0.818 & 0.1058  \\
PCF-ECAPA (C=1024)  & 0.801 & 0.1092 \\
MASV (C=512)  & \textbf{0.805} & \textbf{0.1025}  \\
MASV (C=1024)  & \textbf{0.795} & \textbf{0.0998} \\
\hline
\end{tabular}
\end{table}

\subsection{Evaluation}
We report performance metrics in terms of Equal Error Rate (EER) and minimum Detection Cost Function (minDCF) with \(p_{\text{target}} = 0.01\), \(C_{\text{FA}} = C_{\text{Miss}} = 1\).

Table 1 shows that the MASV model, configured with 512 and 1024 channels, outperforms the baseline ECAPA-TDNN and ResNet models, achieving an average relative improvement of 15.6\% in EER and 15.2\% in minDCF. This confirms the effectiveness of our enhancements in improving speaker verification performance.

The ablation study in Table 2 demonstrates the impact of each component in the MASV architecture. The addition of the LCB-Mamba block reduces EER by 22.7\%, showing its strength in capturing local context without future information. The Tri-Mamba block further enhances performance by integrating both local and global context, while full-scale skip connections improve information flow and stability. The complete MASV model achieves the lowest EER and minDCF, validating the combined contribution of these components to a more robust speaker verification solution.

\begin{table}[tp]
\centering
\caption{Ablation Study of the MASV Architecture}
\begin{tabular}{|c|c|c|c|}
\hline
Configuration & Params & EER  & minDCF\\
\hline
Base Model (512 channels) & 6.2 & 1.158 & 0.1221 \\
+ LCB-Mamba Block & 8.7 & 0.895 & 0.0982 \\
+ Tri-Mamba Block & 9.1 & 0.840 & 0.0947 \\
Complete MASV & 9.2 & \textbf{0.795} & \textbf{0.0875} \\
\hline
\end{tabular}
\end{table}

\begin{figure}[tp]
\centering

\includegraphics[width=0.5\textwidth]{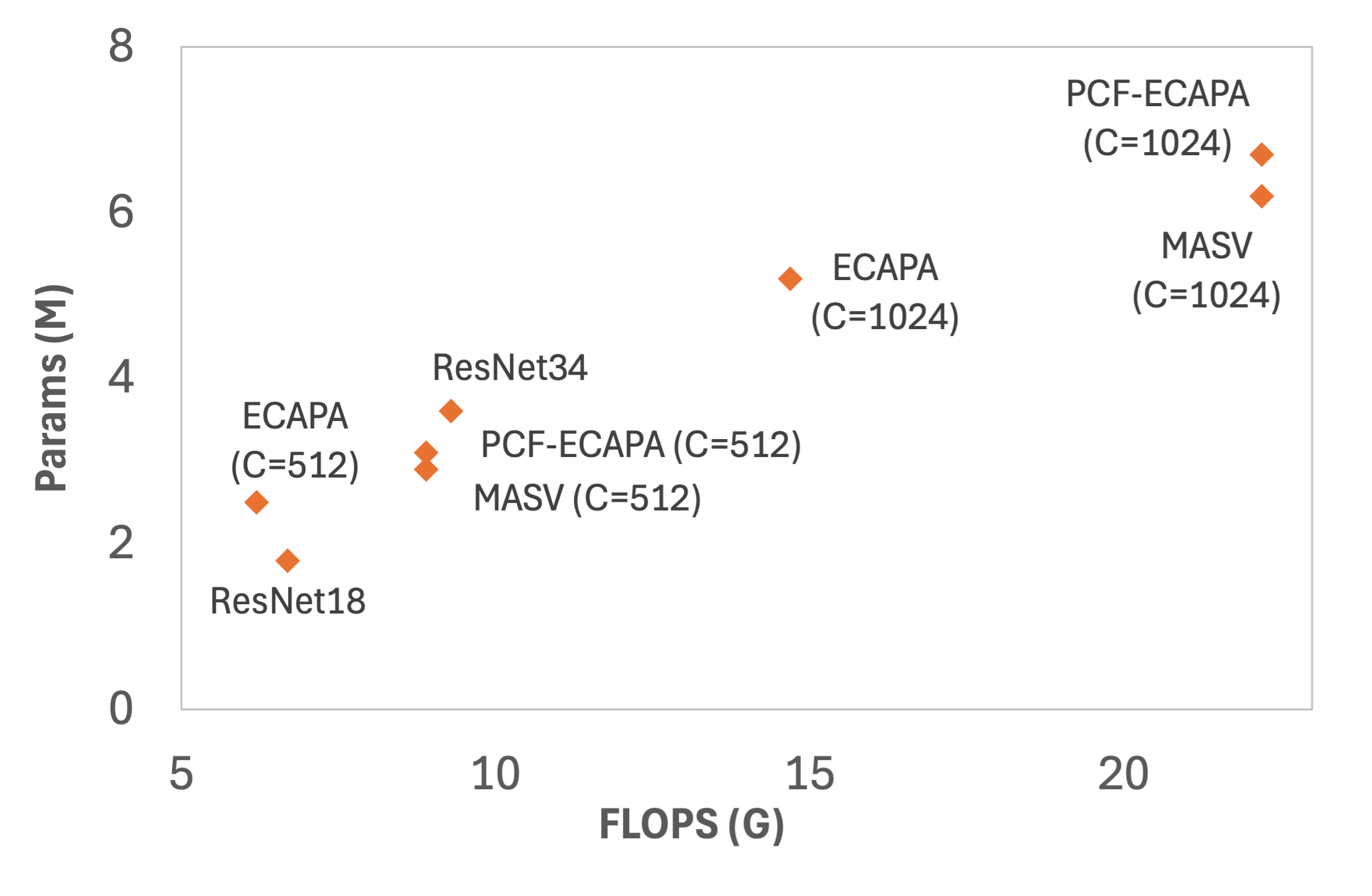}

\caption{Computational Complexity Comparison on}

\label{fig:flop}
\end{figure}

\subsection{Computational Complexity Experiment}

To evaluate the efficiency of the MASV model compared to the baseline models, we analyze their computational complexity by estimating the number of floating-point operations per second (FLOPS) required for each model configuration. Fig. 2 summarizes the number of parameters and FLOPS for each model, including ResNet, ECAPA, PCF-ECAPA, and our proposed MASV.

The MASV model, with both 512 and 1024 channel configurations, demonstrates a balance between parameter size and computational cost. For instance, MASV (C=1024) offers lower FLOPS (6.2G) than PCF-ECAPA (C=1024) while maintaining a competitive parameter count (22.2M), indicating its efficiency in real-time applications.

\section{Conclusion}

This paper introduces the MASV model, a novel speaker verification architecture that enhances real-time accuracy and efficiency by integrating the LCB-Mamba block, Tri-Mamba block, and full-scale skip connections. These innovations address the limitations of existing models by improving local and global context capture and ensuring robust information flow. Our results show that MASV significantly reduces EER and minDCF compared to baseline models while maintaining computational efficiency, making it an effective solution for dynamic, real-world environments.

\bibliographystyle{IEEEbib}
\bibliography{refs}

\begin{thebibliography}{10}

\bibitem{desplanques2020ecapa}
Brecht Desplanques, Jenthe Thienpondt, and Kris Demuynck,
\newblock ``Ecapa-tdnn: Emphasized channel attention, propagation and aggregation in tdnn based speaker verification,''
\newblock {\em arXiv preprint arXiv:2005.07143}, 2020.

\bibitem{zhao2023pcf}
Zhenduo Zhao, Zhuo Li, Wenchao Wang, and Pengyuan Zhang,
\newblock ``Pcf: Ecapa-tdnn with progressive channel fusion for speaker verification,''
\newblock in {\em ICASSP 2023-2023 IEEE International Conference on Acoustics, Speech and Signal Processing (ICASSP)}. IEEE, 2023, pp. 1--5.

\bibitem{liu2024golden}
Tianchi Liu, Kong~Aik Lee, Qiongqiong Wang, and Haizhou Li,
\newblock ``Golden gemini is all you need: Finding the sweet spots for speaker verification,''
\newblock {\em IEEE/ACM Transactions on Audio, Speech, and Language Processing}, 2024.

\bibitem{wang2023vot}
Hongyu Wang, Hui Li, and Bo~Li,
\newblock ``Vot: Revolutionizing speaker verification with memory and attention mechanisms,''
\newblock {\em arXiv preprint arXiv:2312.16826}, 2023.

\bibitem{al2023real}
Khamis~A Al-Karawi,
\newblock ``Real-time adaptive training for forensic speaker verification in reverberation conditions,''
\newblock {\em International Journal of Speech Technology}, vol. 26, no. 4, pp. 1079--1089, 2023.

\bibitem{khan2023battling}
Awais Khan, Khalid~Mahmood Malik, James Ryan, and Mikul Saravanan,
\newblock ``Battling voice spoofing: a review, comparative analysis, and generalizability evaluation of state-of-the-art voice spoofing counter measures,''
\newblock {\em Artificial Intelligence Review}, vol. 56, no. Suppl 1, pp. 513--566, 2023.

\bibitem{gu2021efficiently}
Albert Gu, Karan Goel, and Christopher R{\'e},
\newblock ``Efficiently modeling long sequences with structured state spaces,''
\newblock {\em arXiv preprint arXiv:2111.00396}, 2021.

\bibitem{gu2021combining}
Albert Gu, Isys Johnson, Karan Goel, Khaled Saab, Tri Dao, Atri Rudra, and Christopher R{\'e},
\newblock ``Combining recurrent, convolutional, and continuous-time models with linear state space layers,''
\newblock {\em Advances in neural information processing systems}, vol. 34, pp. 572--585, 2021.

\bibitem{gu2023mamba}
Albert Gu and Tri Dao,
\newblock ``Mamba: Linear-time sequence modeling with selective state spaces,''
\newblock {\em arXiv preprint arXiv:2312.00752}, 2023.

\bibitem{pioro2024moe}
Maciej Pi{\'o}ro, Kamil Ciebiera, Krystian Kr{\'o}l, Jan Ludziejewski, and Sebastian Jaszczur,
\newblock ``Moe-mamba: Efficient selective state space models with mixture of experts,''
\newblock {\em arXiv preprint arXiv:2401.04081}, 2024.

\bibitem{yang2024clinicalmamba}
Zhichao Yang, Avijit Mitra, Sunjae Kwon, and Hong Yu,
\newblock ``Clinicalmamba: A generative clinical language model on longitudinal clinical notes,''
\newblock {\em arXiv preprint arXiv:2403.05795}, 2024.

\bibitem{zhu2024vision}
Lianghui Zhu, Bencheng Liao, Qian Zhang, Xinlong Wang, Wenyu Liu, and Xinggang Wang,
\newblock ``Vision mamba: Efficient visual representation learning with bidirectional state space model,''
\newblock {\em arXiv preprint arXiv:2401.09417}, 2024.

\bibitem{gu2020hippo}
Albert Gu, Tri Dao, Stefano Ermon, Atri Rudra, and Christopher R{\'e},
\newblock ``Hippo: Recurrent memory with optimal polynomial projections,''
\newblock {\em Advances in neural information processing systems}, vol. 33, pp. 1474--1487, 2020.

\bibitem{vaswani2017attention}
Ashish Vaswani, Noam Shazeer, Niki Parmar, Jakob Uszkoreit, Llion Jones, Aidan~N Gomez, {\L}ukasz Kaiser, and Illia Polosukhin,
\newblock ``Attention is all you need,''
\newblock {\em Advances in neural information processing systems}, vol. 30, 2017.

\bibitem{tay2020long}
Yi~Tay, Mostafa Dehghani, Samira Abnar, Yikang Shen, Dara Bahri, Philip Pham, Jinfeng Rao, Liu Yang, Sebastian Ruder, and Donald Metzler,
\newblock ``Long range arena: A benchmark for efficient transformers,''
\newblock {\em arXiv preprint arXiv:2011.04006}, 2020.

\end{thebibliography}

\end{document}